\documentclass[]{spie}

\usepackage{amsmath,amsfonts,amssymb}
\usepackage{graphicx}
\usepackage[colorlinks=true,allcolors=blue]{hyperref}

\title{Design, Fabrication, and Free-Space Characterization of a Two-Photon-Polymerized Nano-printed Photonic Lantern}

\author[a]{Kyra Watts}
\author[a]{Natasha Popenoe}
\author[a]{Benjamin Gerard}
\author[a]{Paul Pax}
\author[a]{Dominic Sanchez}
\author[a]{David Feng}
\author[a]{C. Cole Dunn}
\author[a]{Cari Johnston}
\author[a]{Xiaoxing Xia}

\affil[a]{Lawrence Livermore National Laboratory, 7000 East Avenue,
Livermore, California 94550, USA}

\authorinfo{
Further author information: Send correspondence to Kyra Watts and Dr. Benjamin Gerard\\
E-mail: \texttt{watts30@llnl.gov} and \texttt{gerard3@llnl.gov}
}

\begin{document}
\maketitle

\begin{abstract}
Photonic lanterns provide an interface between multimode optical fields
and arrays of single-mode waveguides, making them promising components
for modal filtering, wavefront sensing, and high-contrast astronomical
instrumentation. We present the design, simulation,
two-photon-polymerization fabrication, and free-space characterization
of a compact photonic lantern optimized for operation at
$ 1050~\mathrm{nm} $. The device was designed to transform a
multimode input into five nominally single-mode output waveguides.
Eigenmode-expansion simulations in Ansys Lumerical MODE were used to
optimize the lantern geometry while minimizing insertion loss and
output-channel power nonuniformity. The optimized structure was
fabricated using a Nanoscribe Photonic Professional GT2
two-photon-polymerization system and developed as a standalone
polymer-waveguide component. The fabricated lantern was characterized
using a free-space optical setup consisting of a $ 635~\mathrm{nm} $
laser for preliminary characterization, collimation and focusing
optics, and camera-based imaging of the transmitted fields.
Measurements were used to assess optical throughput, output intensity
distribution, and light-guiding performance. Differences between
simulated and measured results may arise from the wavelength mismatch,
fabrication tolerances, material and surface-scattering losses,
alignment errors, and deviations from the idealized material and
geometric parameters used in simulation. These results demonstrate a
compact, additively manufactured photonic-lantern platform and
establish a pathway toward free-space-coupled astrophotonic systems
for future high-contrast exoplanet-imaging instrumentation.
\end{abstract}


\section{INTRODUCTION}
\label{sec:introduction}

Photonic lanterns (PLs) provide an interface between a multimode
waveguide and an array of nominally single-mode waveguides. A gradual
transition allows optical power to remain within the guided mode set,
enabling low-loss and reciprocal conversion between the two
representations.\cite{Birks2015} Matching the number of supported
spatial modes at the multimode port to the number of single-mode
channels is a central design requirement for minimizing fundamental
mode-number mismatch loss.

Photonic lanterns have traditionally been fabricated using tapered
fiber bundles, multicore fibers, and ultrafast-laser-inscribed
waveguides\cite{Jovanovic2023}. Two-photon-polymerization-based direct laser writing
(TPP-DLW) provides an alternative approach that supports arbitrary
three-dimensional geometries with submicrometer
resolution.\cite{Maruo2008} The relatively high refractive-index
contrast between the polymer and air permits smaller waveguide cross
sections, tighter bends, and shorter transitions than are generally
practical with weakly guiding fiber structures.

Recent TPP-DLW demonstrations have established the feasibility of
microscale photonic lanterns. Dana et al.\ demonstrated a
$ 375~\mathrm{\mu m} $-long, six-mode, free-standing photonic lantern
printed directly onto a multicore-fiber facet.\cite{Dana2024} Nacar
et al.\ subsequently demonstrated an ultra-compact standalone
$ 1\times5 $ lantern with a sub-$ 100~\mathrm{\mu m} $ multiplexer
region and a total functional length of approximately
$ 350~\mathrm{\mu m} $.\cite{Nacar2026} These results illustrate both
fiber-integrated and detachable standalone implementations of polymer
photonic lanterns.

Scaling photonic lanterns to larger mode counts has also been
investigated for space-division multiplexing.\cite{VelazquezBenitez2018}
Efficient numerical methods for tapered photonic structures are
important in this context because full three-dimensional simulations
can become computationally demanding as device length and mode count
increase.\cite{Tschernig2025} More recent work has applied 3D-printed
photonic lanterns to massive-scale spatial multiplexing of multimode
vertical-cavity surface-emitting lasers, demonstrating the broader
integration potential of the fabrication platform.\cite{DanaVCSEL2026}

In astrophotonics, photonic lanterns have been proposed for modal
filtering, spectroscopy, coherent imaging, and focal-plane wavefront
sensing. In particular, the intensity distribution across the
single-mode outputs can encode information about the incident
wavefront, providing a potential route to wavefront sensing and
control for high-contrast exoplanet imaging.\cite{LinExoplanet}

In this work, we present the design, simulation, TPP fabrication, and
free-space characterization of a compact five-output photonic lantern
optimized for operation at $ 1050~\mathrm{nm} $. Eigenmode-expansion
simulations were used to optimize the transition length and geometric
parameters while limiting insertion loss and output-power
nonuniformity. The optimized device was fabricated using a Nanoscribe
Photonic Professional GT2 system and characterized with a free-space
optical setup. The measurements assess optical throughput, output
intensity distribution, and light-guiding performance, providing a
basis for future free-space-coupled astrophotonic instrumentation.

\section{DEVICE DESIGN AND SIMULATION}
\label{sec:design}

\subsection{Photonic-lantern architecture and geometry}

The photonic lantern was adapted from the designs reported by Dana
et al.~\cite{Dana2024} and Nacar et al.~\cite{Nacar2026} for
free-space characterization. The device comprises a multimode (MM)
input taper, a compact mode-multiplexing (MUX) transition, and five
nominally single-mode (SM) output channels. The MM taper transforms
the $ 29.9~\mathrm{\mu m} $ free-space-facing aperture into a
$ 2.0~\mathrm{\mu m} $ interface at the MUX. Within the MUX, the
composite MM cross section evolves into five spatially separated
waveguides. The output tapers subsequently expand the individual
guided fields to facilitate free-space collection and imaging.

The MUX geometry consists of one central waveguide and four outer
waveguides. Each waveguide has a circular cross section with a radius
of $ 0.5~\mathrm{\mu m} $. The central waveguide remains on the
optical axis, whereas the outer waveguides occupy the four coordinate
combinations
\begin{equation}
(y,z)=
\left\{
\bigl(c(x),c(x)\bigr),
\bigl(c(x),-c(x)\bigr),
\bigl(-c(x),c(x)\bigr),
\bigl(-c(x),-c(x)\bigr)
\right\}.
\end{equation}

The coordinate magnitude $ c(x) $ follows a normalized sigmoid:
\begin{align}
f(x)
&=
\frac{1}{1+\exp\left[-(a/L_{\mathrm{MUX}})x\right]}, \\
\widehat{f}(x)
&=
\frac{f(x)-f(x_0)}{f(x_1)-f(x_0)}, \\
c(x)
&=
c_{\mathrm{MM}}
+
\left(c_{\mathrm{SM}}-c_{\mathrm{MM}}\right)\widehat{f}(x),
\end{align}
where $ x_0=-L_{\mathrm{MUX}}/2 $, $ x_1=L_{\mathrm{MUX}}/2 $,
$ L_{\mathrm{MUX}}=77~\mathrm{\mu m} $, $ a=2.5 $,
$ c_{\mathrm{MM}}=0.5~\mathrm{\mu m} $, and
$ c_{\mathrm{SM}}=3.7~\mathrm{\mu m} $ (radial separation). The quantity $ c(x) $ is the
magnitude of each transverse coordinate and is not the radial
center-to-center separation. The radial separation between the
central waveguide and each outer waveguide is $ \sqrt{2}c(x) $.

The MUX geometry was represented by 154 cylindrical sections with a
nominal longitudinal thickness of $ 0.5~\mathrm{\mu m} $, resulting
in a total length of $ 77~\mathrm{\mu m} $. The MM input taper was
similarly discretized into cylindrical sections. Its diameter varied
linearly from $ 29.9~\mathrm{\mu m} $ at the input to
$ 2.0~\mathrm{\mu m} $ at the interface with the MUX, over a total
length of $ 300~\mathrm{\mu m} $. Similarly the SM output tapers varied lengths from $ 1~\mathrm{\mu m} $ to match the MUX arms to $ 13.4~\mathrm{\mu m} $ at the outputs, over a length of $ 200~\mathrm{\mu m} $.

The taper diameters were selected
for free space characterization with an input beam with an $f/10$ focal ratio. Using
\begin{equation}
\mathrm{NA} = \frac{1}{2f/\#} = \frac{1}{2(10)} = 0.05,
\end{equation}
the mode-field diameter was calculated at a wavelength of
$\lambda = 1050~\mathrm{nm}$ according to
\begin{equation}
d = \frac{2\lambda\sqrt{N}}{\pi\,\mathrm{NA}},
\end{equation}
where $N$ is the number of guided modes. For the MM input,
$N=5$, giving $d \approx 29.9~\mathrm{\mu m}$, while $N=1$ at the
SM end gives $d \approx 13.4~\mathrm{\mu m}$.

\begin{figure}[ht]
    \centering
    \includegraphics[width=\textwidth]{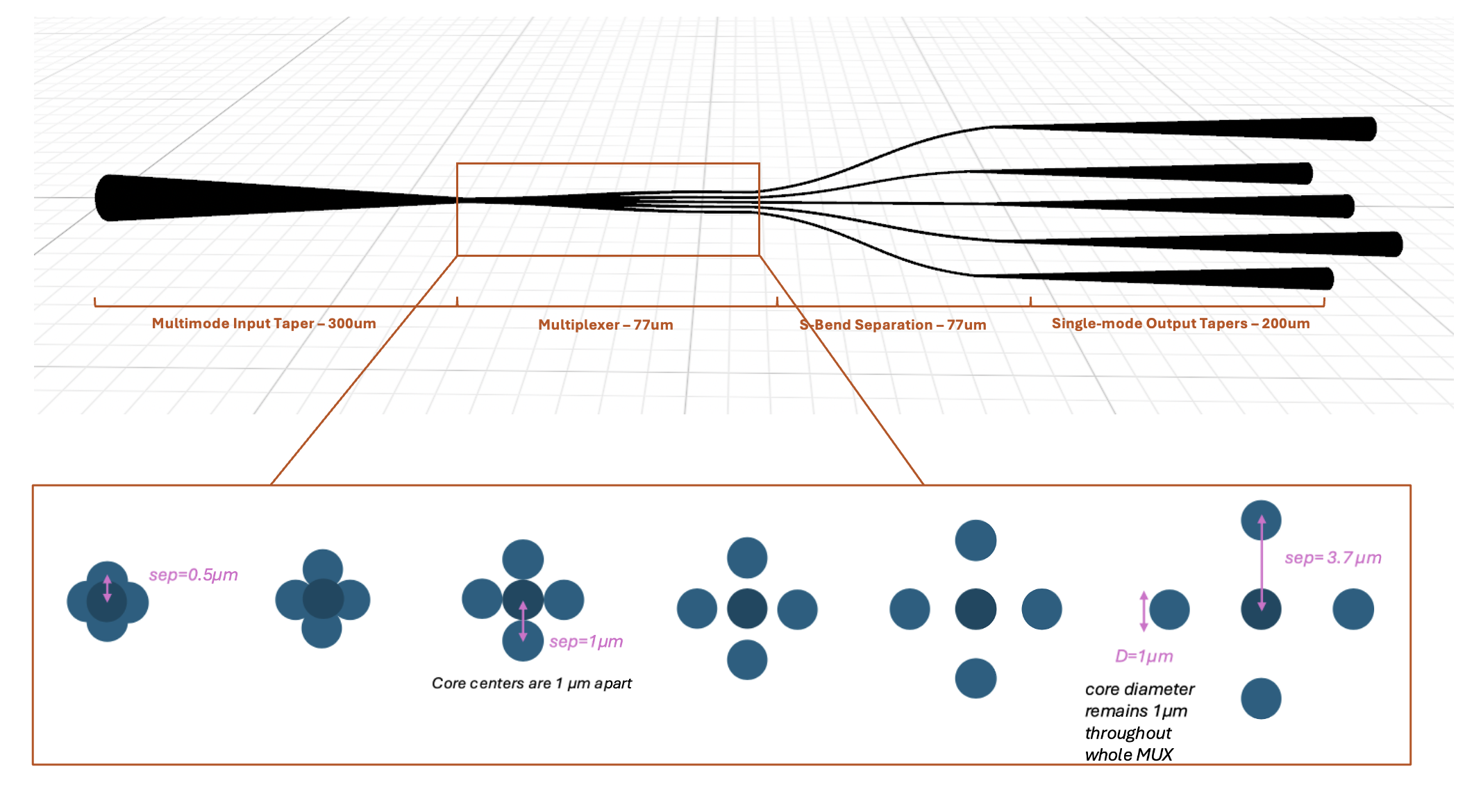}
    \caption{Schematic of the five-channel photonic lantern. The
    multimode input taper feeds the MUX region, within which the
    composite MM waveguide evolves into five spatially separated
    output channels. Each channel subsequently expands through an
    output taper. The transverse evolution of the MUX cores is shown
    beneath the complete device. The nominal device dimensions are
    summarized in Table~\ref{tab:design_parameters}.}
    \label{fig:device_schematic}
\end{figure}

Figure~\ref{fig:device_schematic} illustrates the complete device,
including the MM input taper, MUX transition, and five output
waveguides.

\begin{table}[ht]
    \caption{Nominal photonic-lantern design parameters.}
    \label{tab:design_parameters}
    \centering
    \begin{tabular}{|l|c|}
        \hline
        Parameter & Value \\
        \hline
        Design wavelength
            & $ 1050~\mathrm{nm} $ \\
        Free-space-facing MM taper diameter
            & $ 29.9~\mathrm{\mu m} $ \\
        Diameter at the taper--MUX interface
            & $ 2.0~\mathrm{\mu m} $ \\
        MUX length
            & $ 77~\mathrm{\mu m} $ \\
        MUX waveguide diameter
            & $ 1.0~\mathrm{\mu m} $ \\
        MUX Arm Sep at MM end
            & $ 0.5~\mathrm{\mu m} $ \\
        MUX Arm Sep at SM end
            & $ 3.7~\mathrm{\mu m} $ \\
        Final SM output-taper diameter
            & $ 13.4~\mathrm{\mu m} $ \\
        Nominal final output-arm spacing
            & $ 27.38~\mathrm{\mu m} $ \\
        Total device length
            & $ 654~\mathrm{\mu m} $ \\
        Waveguide-core refractive index
            & $ 1.53 $ \\
        Surrounding-medium refractive index
            & $ 1.00 $ \\
        \hline
    \end{tabular}
\end{table}

\subsection{Eigenmode-expansion models}

The MM taper and MUX were modeled independently using the
three-dimensional eigenmode-expansion (EME) solver in Ansys Lumerical
MODE Solutions. The nominal wavelength was
$ \lambda_0=1.05~\mathrm{\mu m} $, and propagation was defined along
the $ x $-axis. A real refractive index of $ 1.53 $ was assigned to
the waveguide cores, and the nominal surrounding-medium index was
$ 1.00 $.

The principal settings used in the simulations are summarized in
Table~\ref{tab:eme_settings}. The number of modes in the EME
cell-group basis is distinct from the number of modes retained at the
external ports. The larger cell-group basis represents the field
evolution within the device, whereas the port basis defines the modal
channels retained in the scattering matrix.

\begin{table}[ht]
    \caption{Principal EME simulation settings.}
    \label{tab:eme_settings}
    \centering
    \begin{tabular}{|l|c|c|}
        \hline
        Parameter & MM taper & MUX \\
        \hline
        Solver
            & 3D EME & 3D EME \\
        Propagation direction
            & $ x $-axis & $ x $-axis \\
        Wavelength
            & $ 1.05~\mathrm{\mu m} $
            & $ 1.05~\mathrm{\mu m} $ \\
        Cell-group basis modes
            & 100 & 30 \\
        Longitudinal EME cells
            & 100 & 30 \\
        MM-side port modes
            & 10 & 10 \\
        SM-side port modes
            & Not applicable & 2 per SM port \\
        Total channels on each side
            & 10 & 10 \\
        Scattering-matrix size
            & $ 20\times20 $ & $ 20\times20 $ \\
        Transverse boundaries
            & PML & PML \\
        \hline
    \end{tabular}
\end{table}

For the taper model, perfectly matched layer (PML) boundary conditions
were applied at the minimum and maximum $ y $ and $ z $ boundaries.
The full transverse simulation span was
$ 39.9~\mathrm{\mu m}\times39.9~\mathrm{\mu m} $. The modal window at
the $ 2.0~\mathrm{\mu m} $ end was
$ 7.0~\mathrm{\mu m}\times7.0~\mathrm{\mu m} $, whereas the window at
the $ 29.9~\mathrm{\mu m} $ end used the full transverse span. These
restricted port windows reduced the inclusion of distant radiation
and PML-associated modes in the selected port basis.

Unlike finite-difference methods, EME does not employ a uniform
longitudinal mesh. Instead, the structure is divided into longitudinal
cells, and the field within each cell is expanded in the local
eigenmode basis. The taper model used 100 cells, while the MUX
calculation used 30 longitudinal cells and 30 cell-group modes.
These parameters were chosen to meet numerical convergence requirements, as described further in Sec. 2.5.

\begin{figure}[ht]
    \centering
    \includegraphics[width=\textwidth]{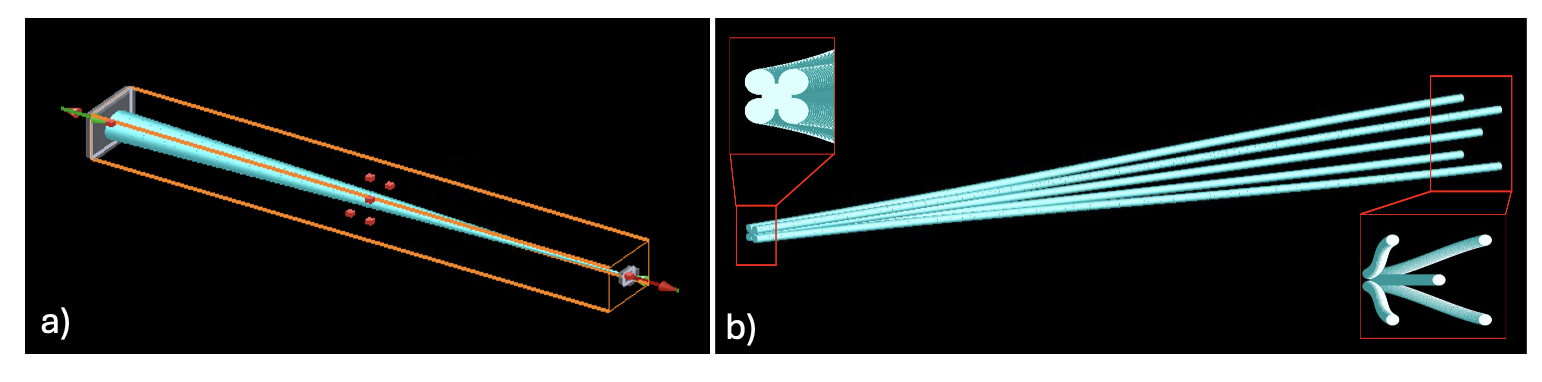}
    \caption{Simulation geometries for (a) the input taper with its
    surrounding EME region and ports and (b) the MUX. Both geometries
    were constructed using cylindrical sections stacked along the
    propagation axis.}
    \label{fig:geom_models}
\end{figure}

Figure~\ref{fig:geom_models} illustrates the discrete geometric
models constructed in Lumerical and used in the EME simulations of
the tapers and MUX.

\subsection{Modal excitation and polarization}

Each calculation used the forward-propagating, power-normalized
eigenmodes of the corresponding input port. A single eigenmode was
excited during each launch, and the transmitted field was projected
onto the selected output channels. Repeating this procedure for all
input channels produced the complete mode-resolved transmission
matrix.

The port fields were calculated as full-vector eigenmodes of the
waveguide cross section. Polarization was therefore determined from
the electric-field distribution of each eigenmode rather than imposed
as a plane-wave polarization. At each nominally SM port, two modes
were retained to represent the two orthogonal polarization channels.
TE-like and TM-like labels were assigned after verifying the dominant
electric-field components of the calculated mode profiles.

In the multimode-taper simulations, transmission was evaluated from
the $ 2.0~\mathrm{\mu m} $ end toward the
$ 29.9~\mathrm{\mu m} $ end. Similarly, the MUX-analysis script
extracted transmission from the five SM-side ports toward the MM-side
port. These directions are opposite to the nominal free-space-input
direction of the assembled lantern. Because the models used linear,
isotropic, reciprocal materials, the forward- and reverse-transmission
matrices are related by reciprocity. However, launch-specific
insertion loss can depend on the chosen modal basis and launch
weighting, so the propagation direction and associated S-matrix block
are stated explicitly.

\begin{figure}[ht]
    \centering
    \includegraphics[width=\textwidth]{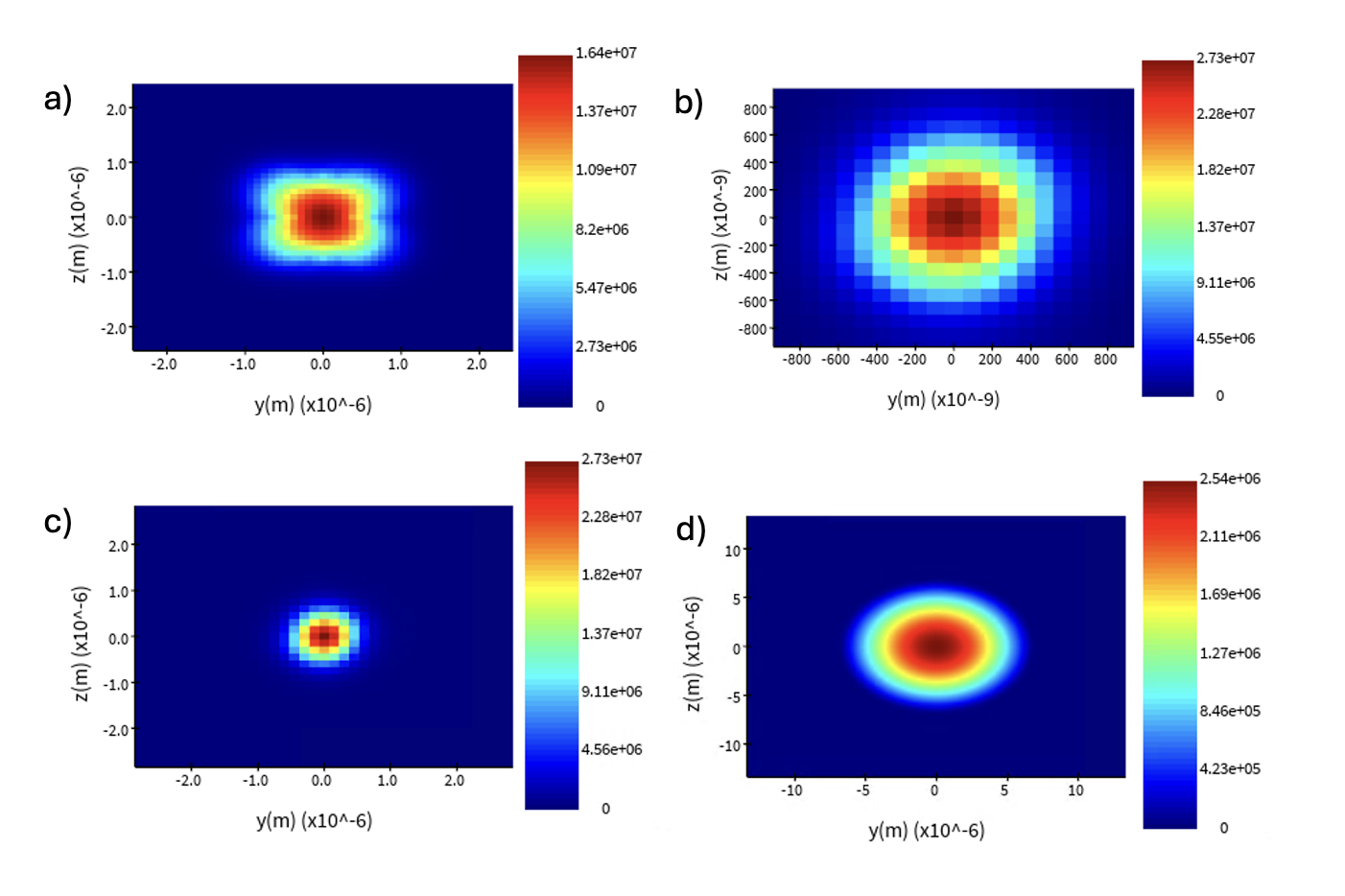}
    \caption{Representative simulated mode profiles for the MUX and
    SM output-taper models.}
    \label{fig:mode_profiles}
\end{figure}

Figure~\ref{fig:mode_profiles} illustrates representative mode
profiles at (a) the input of the MUX, (b) the output of one SM MUX
arm, (c) the input of an SM taper, and (d) the output of an SM taper.

\subsection{Scattering-matrix formulation and loss metrics}

The scattering matrix, or S-matrix, provides a mode-resolved
description of reflection, transmission, and mode conversion within
the photonic lantern. For a two-sided optical device, the incident
modal-amplitude vectors, $ \mathbf{a}_1 $ and $ \mathbf{a}_2 $, and
the outgoing modal-amplitude vectors, $ \mathbf{b}_1 $ and
$ \mathbf{b}_2 $, are related by
\begin{equation}
    \begin{bmatrix}
        \mathbf{b}_1 \\
        \mathbf{b}_2
    \end{bmatrix}
    =
    \begin{bmatrix}
        \mathbf{S}_{11} & \mathbf{S}_{12} \\
        \mathbf{S}_{21} & \mathbf{S}_{22}
    \end{bmatrix}
    \begin{bmatrix}
        \mathbf{a}_1 \\
        \mathbf{a}_2
    \end{bmatrix}.
    \label{eq:s_matrix}
\end{equation}

The submatrices $ \mathbf{S}_{11} $ and $ \mathbf{S}_{22} $ describe
reflection at sides 1 and 2, respectively, whereas
$ \mathbf{S}_{21} $ and $ \mathbf{S}_{12} $ describe transmission in
the two propagation directions. Each S-matrix element is a complex
quantity containing both amplitude and phase information. For
power-normalized modes, $ |S_{pq,mj}|^2 $ is the fraction of power
coupled from incident mode $ j $ on side $ q $ to outgoing mode $ m $
on side $ p $.

Ten modal channels were retained on each side of the MUX, resulting in
a $ 20\times20 $ S-matrix composed of four $ 10\times10 $ blocks. The
multimode side contained ten selected eigenmodes, while the
single-mode side contained two polarization channels for each of the
five waveguides. Within a transmission block, diagonal elements
represent coupling between correspondingly ordered channels, while
off-diagonal elements describe coupling between differently ordered
channels.

Let $ \mathbf{H} $ denote the $ 10\times10 $ transmission-amplitude
block associated with the propagation direction being analyzed. Thus,
$ \mathbf{H}=\mathbf{S}_{21} $ for excitation from side 1 and
$ \mathbf{H}=\mathbf{S}_{12} $ for excitation from side 2. For input
mode $ j $, the captured transmitted power is
\begin{equation}
    T_j
    =
    \sum_{m=1}^{N_{\mathrm{out}}}
    \left|H_{mj}\right|^2,
    \qquad
    N_{\mathrm{out}}=10.
    \label{eq:transmitted_power}
\end{equation}

The corresponding insertion loss is
\begin{equation}
    \mathrm{IL}_j
    =
    -10\log_{10}(T_j),
\end{equation}
and the mean launch-dependent insertion loss is
\begin{equation}
    \overline{\mathrm{IL}}
    =
    \frac{1}{N_{\mathrm{in}}}
    \sum_{j=1}^{N_{\mathrm{in}}}\mathrm{IL}_j,
    \qquad
    N_{\mathrm{in}}=10.
    \label{eq:mean_insertion_loss}
\end{equation}

Let $ \mathbf{G} $ denote the reflection block associated with the
excited side, such that $ \mathbf{G}=\mathbf{S}_{11} $ for excitation
from side 1 and $ \mathbf{G}=\mathbf{S}_{22} $ for excitation from
side 2. The reflected power associated with input mode $ j $ is
\begin{equation}
    R_j
    =
    \sum_{m=1}^{N_{\mathrm{in}}}
    \left|G_{mj}\right|^2.
\end{equation}

For a unit-power modal excitation, the residual power not represented
by the selected transmitted and reflected port modes is
\begin{equation}
    P_{\mathrm{uncaptured},j}
    =
    1-T_j-R_j.
\end{equation}

This residual can include radiation absorbed by the PML boundaries,
coupling to modes outside the selected port basis, and numerical
truncation error. Because real refractive indices were used in the
model, material absorption was not included.

For the SM-side launch convention used in the MUX script, the two
polarization channels associated with each of the five SM ports were
grouped using alternating modal indices:
\begin{equation}
    \mathcal{J}_1=\{1,3,5,7,9\},
    \qquad
    \mathcal{J}_2=\{2,4,6,8,10\}.
\end{equation}

The mean transmission for polarization group $ p $ was calculated as
\begin{equation}
    \overline{T}_p
    =
    \frac{1}{5}
    \sum_{j\in\mathcal{J}_p}T_j,
    \qquad
    p\in\{1,2\}.
\end{equation}

The polarization-dependent loss was then calculated as
\begin{equation}
    \mathrm{PDL}
    =
    10\log_{10}
    \left[
        \frac{\max\left(\overline{T}_1,\overline{T}_2\right)}
             {\min\left(\overline{T}_1,\overline{T}_2\right)}
    \right].
\end{equation}

For completeness, conventional mode-dependent loss is defined using
the power-transmission eigenvalues $ \lambda_k $ of
$ \mathbf{H}^{\dagger}\mathbf{H} $, which are equal to the squared
singular values of $ \mathbf{H} $:
\begin{equation}
    \mathrm{MDL}
    =
    10\log_{10}
    \left(
        \frac{\lambda_{\max}}{\lambda_{\min}}
    \right).
    \label{eq:mdl}
\end{equation}

This definition is invariant under unitary transformations of the
input and output modal bases. Conventional MDL was not calculated for
the results reported below because the implemented analysis used
row-summed output powers rather than the eigenvalues of
$ \mathbf{H}^{\dagger}\mathbf{H} $.

The row-summed output powers were calculated as
\begin{equation}
    q_m
    =
    \sum_{j=1}^{N_{\mathrm{in}}}
    \left|H_{mj}\right|^2.
\end{equation}

The output-power nonuniformity was then defined as
\begin{equation}
    \Delta_{\mathrm{out}}
    =
    10\log_{10}
    \left(
        \frac{\max_m q_m}{\min_m q_m}
    \right).
\end{equation}

This quantity measures the variation in power delivered to the
selected output modes under equal, mutually incoherent excitation of
the selected input channels. It is not equivalent to conventional MDL
in Eq.~\eqref{eq:mdl}. All modal-nonuniformity values reported in this
work were calculated using $ \Delta_{\mathrm{out}} $ and are therefore
identified as output-power nonuniformity rather than MDL.

The same S-matrix framework was applied independently to the MM input
taper and the five SM output tapers. For the input taper, ten modes
were retained at both ports, giving a $ 10\times10 $ transmission
matrix and a $ 20\times20 $ complete S-matrix. Each of the ten input
modes was launched independently. The resulting $ T_j $,
$ \mathrm{IL}_j $, mean insertion loss, and output-power nonuniformity
$ \Delta_{\mathrm{out}} $ were calculated using the definitions
above. The corresponding reflection and uncaptured-power fractions
were also evaluated for each modal launch.

Each SM output taper was analyzed using the two orthogonally polarized
fundamental modes, identified as the fundamental TE-like and TM-like
modes. Consequently, each output-taper model produced a
$ 2\times2 $ transmission matrix and a $ 4\times4 $ complete
S-matrix. The polarization-resolved transmitted powers were
\begin{align}
    T_{\mathrm{TE}}
    &=
    \sum_{m=1}^{2}
    \left|H_{m,\mathrm{TE}}\right|^2, \\
    T_{\mathrm{TM}}
    &=
    \sum_{m=1}^{2}
    \left|H_{m,\mathrm{TM}}\right|^2.
\end{align}

The corresponding insertion losses were
\begin{align}
    \mathrm{IL}_{\mathrm{TE}}
    &=
    -10\log_{10}\left(T_{\mathrm{TE}}\right), \\
    \mathrm{IL}_{\mathrm{TM}}
    &=
    -10\log_{10}\left(T_{\mathrm{TM}}\right).
\end{align}

The mean insertion loss of an output taper was calculated as
\begin{equation}
    \overline{\mathrm{IL}}_{\mathrm{SM}}
    =
    \frac{
        \mathrm{IL}_{\mathrm{TE}}
        +
        \mathrm{IL}_{\mathrm{TM}}
    }{2},
\end{equation}
and its polarization-dependent loss was evaluated as
\begin{equation}
    \mathrm{PDL}_{\mathrm{SM}}
    =
    10\log_{10}
    \left[
        \frac{\max\left(T_{\mathrm{TE}},T_{\mathrm{TM}}\right)}
             {\min\left(T_{\mathrm{TE}},T_{\mathrm{TM}}\right)}
    \right].
\end{equation}

For each output taper, the reported modal-uniformity metric was
calculated from the row-summed powers of its $ 2\times2 $ transmission
matrix. Conventional MDL would instead require the eigenvalues of
$ \mathbf{H}^{\dagger}\mathbf{H} $ and was not used for the reported
results. This component-wise analysis enabled the transmission
performance of the input taper, MUX, and output tapers to be assessed
using a consistent set of metrics while retaining the modal basis
appropriate to each component.

\begin{figure}[ht]
    \centering
    \includegraphics[width=0.6\textwidth]{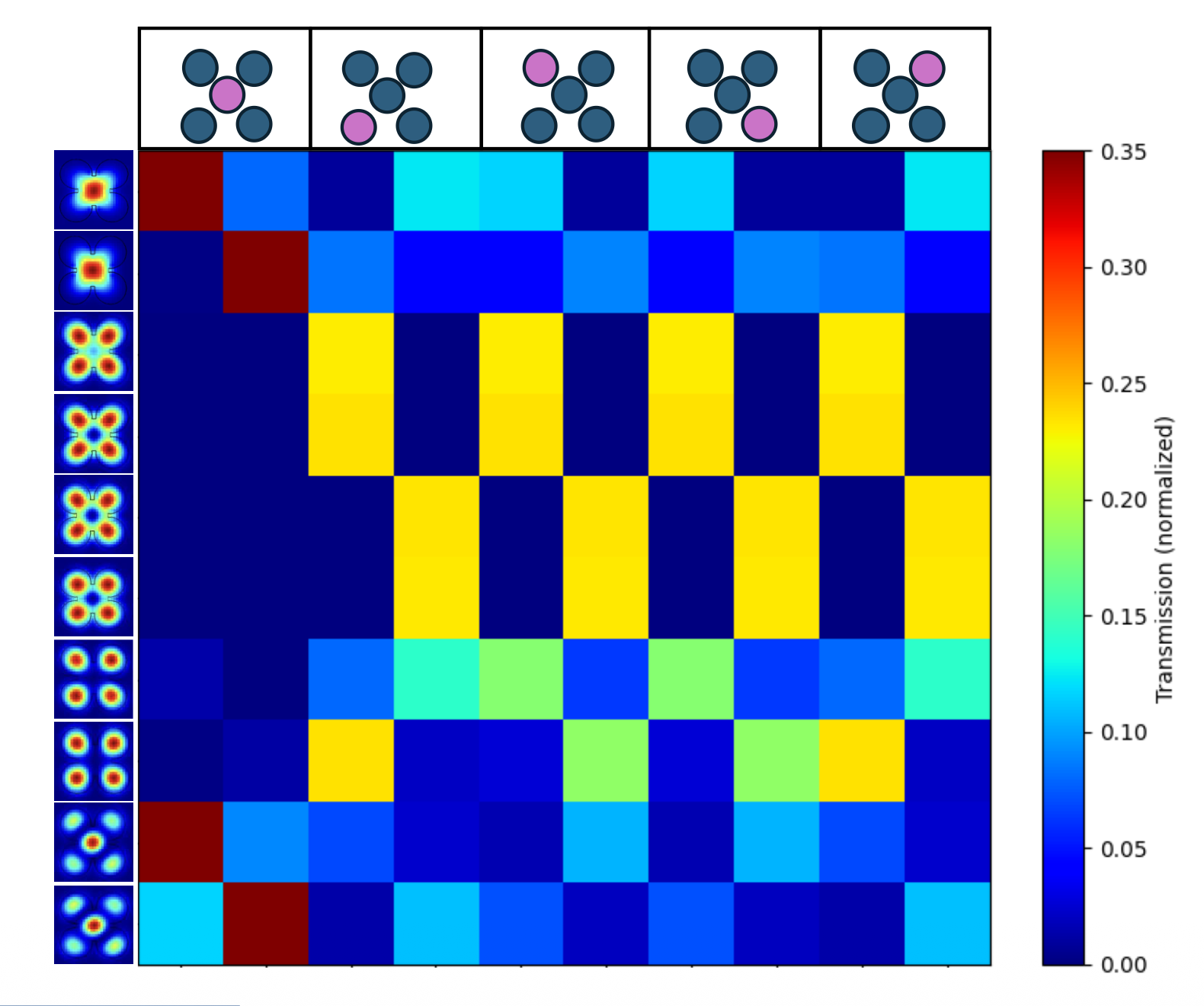}
    \caption{Magnitude-squared S-matrix, $ |S_{mj}|^2 $, for the
    finalized MUX design. Each element represents the fraction of
    power coupled from incident modal channel $ j $ to outgoing modal
    channel $ m $.}
    \label{fig:matrix}
\end{figure}

Figure~\ref{fig:matrix} shows the magnitude-squared S-matrix obtained
for the finalized MUX design. The rows represent outgoing modal
channels, while the columns represent incident modal channels. The
four quadrants correspond to the two reflection and two transmission
blocks defined in Eq.~\eqref{eq:s_matrix}.

\subsection{Design constraints and numerical convergence}

The MUX and taper geometries were parameterized and evaluated
independently. The MUX parameters included its length, sigmoid
coefficient, core radius, and initial and final transverse
coordinates. The taper parameters included its endpoint diameters,
transition length, and taper exponent. Candidate geometries were
evaluated using insertion loss and output-power nonuniformity while
enforcing a minimum transverse feature size of
$ 1.0~\mathrm{\mu m} $, consistent with the nominal Nanoscribe
fabrication constraint.

Numerical convergence was assessed by increasing both the number of
longitudinal EME cells and the number of cell-group basis modes until
the reported loss metrics stabilized. An example convergence plot for a MM taper design is shown below in Fig. ~\ref{fig:convergence}

\begin{figure}[ht]
    \centering
    \includegraphics[width=0.6\textwidth]{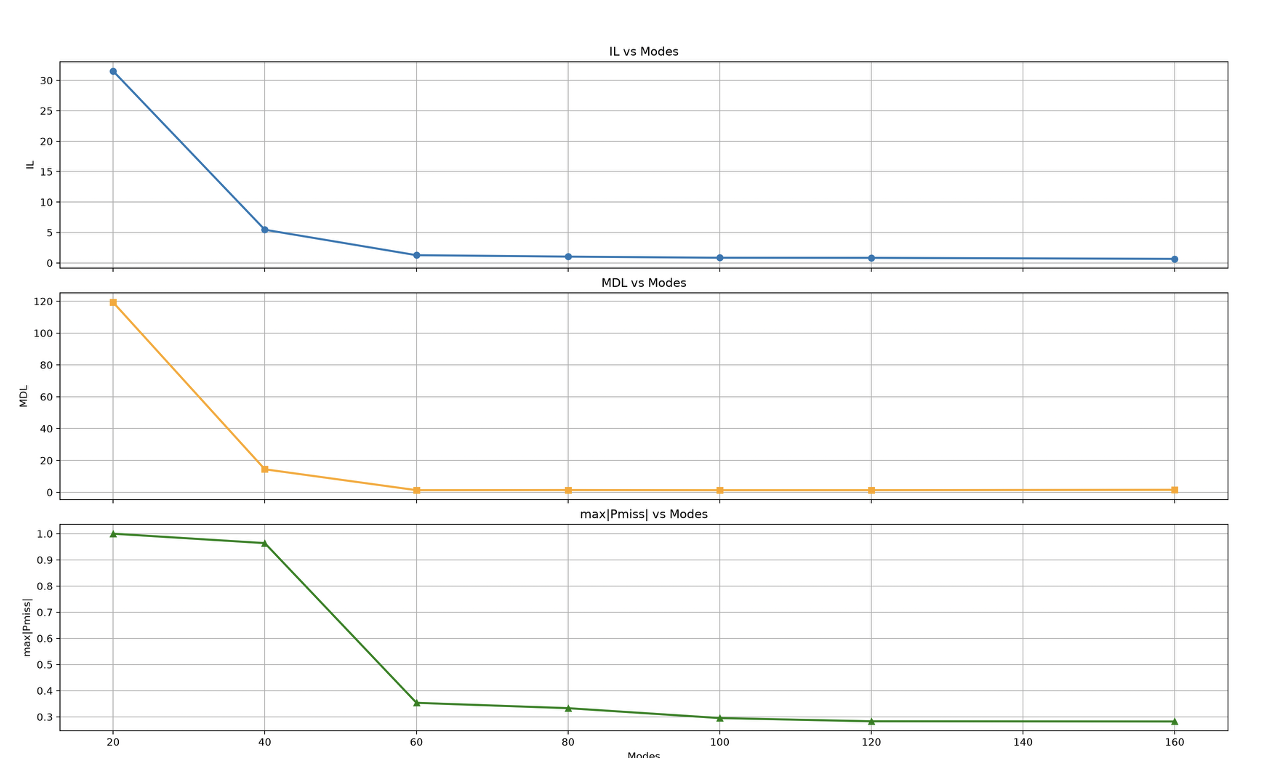}
    \caption{An example of a convergence plot for one of the MM taper designs that was simulated}
    \label{fig:convergence}
\end{figure}

The single-mode taper was evaluated using an approach similar to that
used for the multimode taper. However, the single-mode-taper model was
less computationally expensive because of its smaller geometry and
the smaller number of evaluated modes, consisting of the fundamental
TE-like and TM-like modes.

\subsection{Simulation results}

\begin{figure}[ht]
    \centering
    \includegraphics[width=0.6\textwidth]{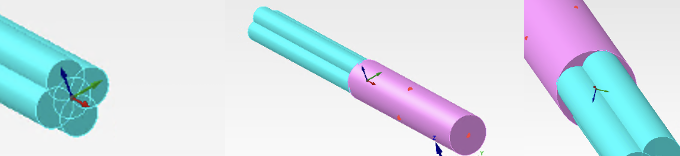}
    \caption{An image of the simulated circular-taper to flower-MUX interface.}
    \label{fig:interface}
\end{figure}

The final simulated mean insertion-loss and output-power-nonuniformity
values are summarized in Table~\ref{tab:losses}. The interface between
the MM input taper and the MUX was also simulated to assess the coupling
loss associated with transforming the circular MM mode into the
inscribed MUX geometry, as shown in Fig~\ref{fig:interface}. The resulting interface insertion loss was
less than $0.4~\mathrm{dB}$ and is therefore small compared with the
other component losses, although it is included in the total
insertion-loss estimate below.

\begin{table}[ht]
    \label{tab:losses}
    \centering
    \begin{tabular}{|l|c|c|}
        \hline
        Component & Mean IL (dB) & $\Delta_{\mathrm{out}}$ (dB) \\
        \hline
        Multimode input taper
            & 3.00949 & 3.56303 \\
        MM--MUX interface
            & $<0.4$ & -- \\
        MUX
            & 0.56000 & 0.86000 \\
        S-bend separator
            & 0.56000 & 0.86000 \\
        Single-mode output taper
            & 1.98347 & 0.50660 \\
        \hline
    \end{tabular}
    \caption{Simulated mean insertion loss and output-power
    nonuniformity of each photonic-lantern component at
    $1050~\mathrm{nm}$. The S-bend separator was not simulated
    independently and was assigned the same loss and nonuniformity as
    the MUX.}
\end{table}

For cascaded components whose power-transmission efficiencies can be
treated independently, the linear transmission factors multiply and
the corresponding insertion losses add in decibels, consistent with
standard optical-link power-budget analysis~\cite{Keiser2013}. Thus,
the estimated insertion loss of the complete photonic lantern is
\begin{align}
    \mathrm{IL}_{\mathrm{total}}
    &=
    \mathrm{IL}_{\mathrm{MM}}
    +
    \mathrm{IL}_{\mathrm{interface}}
    +
    \mathrm{IL}_{\mathrm{MUX}}
    +
    \mathrm{IL}_{\mathrm{S\text{-}bend}}
    +
    \mathrm{IL}_{\mathrm{SM}} \\
    &=
    3.00949 + \mathrm{IL}_{\mathrm{interface}}
    + 0.56000 + 0.56000 + 1.98347 \\
    &<
    6.51296~\mathrm{dB},
\end{align}
where $\mathrm{IL}_{\mathrm{interface}}<0.4~\mathrm{dB}$. If the
interface loss is neglected, the estimate is
\begin{equation}
    \mathrm{IL}_{\mathrm{total}}
    \approx 6.11296~\mathrm{dB}.
\end{equation}

The S-bend separator was not simulated as a separate component.
Because its cross-sectional geometry and characteristic aspect ratio
were assumed to be comparable to those of the MUX, its insertion loss
and output-power nonuniformity were conservatively taken to be equal
to the corresponding MUX values. This approximation should be verified
with a dedicated S-bend simulation in future work. The total estimate
also assumes negligible intercomponent reflections, interference, and
additional coupling losses beyond the simulated MM--MUX interface.
Only one SM output-taper loss is included because each optical path
passes through one of the five output tapers.

\subsection{STL-based geometry model}

\begin{figure}[ht]
    \centering
    \includegraphics[width=0.6\textwidth]{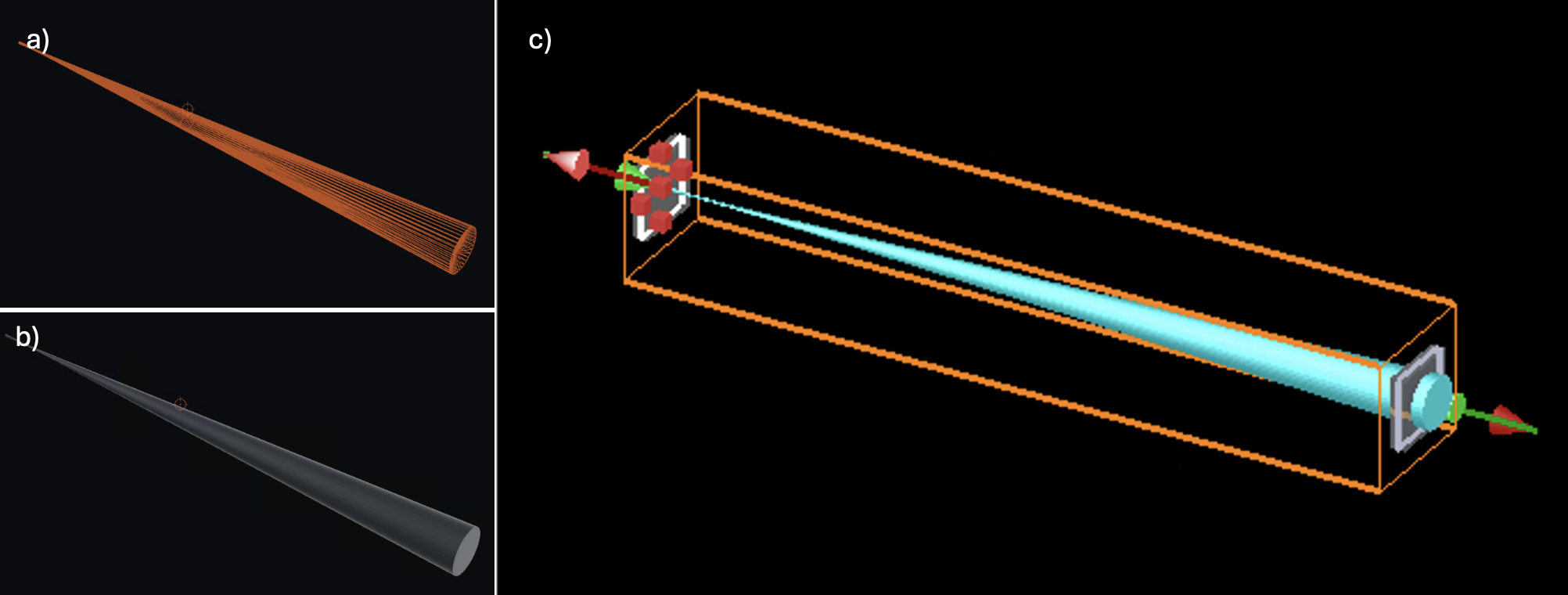}
    \caption{STL-based output-taper model: (a) wireframe 3D model,
    (b) solid 3D model, and (c) the imported geometry in Ansys
    Lumerical MODE.}
    \label{fig:stl}
\end{figure}

An alternative model was investigated using an STL representation of
the photonic-lantern geometry for the single-mode taper. The STL file
was imported into Ansys Lumerical MODE, assigned the appropriate
refractive index, and aligned with the $ x $-axis as the propagation
direction. A three-dimensional EME region was then configured around
the imported structure using the same wavelength, boundary
conditions, port definitions, and modal basis as the parameterized
model, as shown in Fig.~\ref{fig:stl}.

The STL representation produced a lower simulated insertion loss than
the discretized geometry. This improvement was attributed to the more
continuous representation of the longitudinal transition, which
reduced artificial discontinuities between adjacent geometric
sections. The insertion loss decreased from
$ 1.98347~\mathrm{dB} $ to $ 0.617101~\mathrm{dB} $. Before this
result was used quantitatively, convergence with respect to the EME
cell count was verified.

As a qualitative estimate, the MM input taper may exhibit a similar
improvement if represented using a continuous STL geometry. The STL
model reduced the SM taper insertion loss by
\begin{equation}
    \Delta\mathrm{IL}_{\mathrm{STL}}
    =
    1.98347-0.617101
    =
    1.366369~\mathrm{dB}.
\end{equation}

Assuming that the MM taper experiences the same reduction in
insertion loss, its estimated loss would decrease from
$ 3.00949~\mathrm{dB} $ to
\begin{equation}
    \mathrm{IL}_{\mathrm{MM,STL}}
    =
    3.00949-1.366369
    =
    1.643121~\mathrm{dB}.
\end{equation}

Equivalently, this assumption applies the same multiplicative
improvement in power transmission to both tapers. Using the
STL-based estimates for the MM and SM tapers, retaining the simulated
MUX loss, including the estimated MM--MUX interface loss, and assigning
the S-bend separator the same loss as the MUX, the total device
insertion loss is estimated as
\begin{align}
    \mathrm{IL}_{\mathrm{device,STL}}
    &=
    \mathrm{IL}_{\mathrm{MM,STL}}
    +
    \mathrm{IL}_{\mathrm{interface}}
    +
    \mathrm{IL}_{\mathrm{MUX}}
    +
    \mathrm{IL}_{\mathrm{S\text{-}bend}}
    +
    \mathrm{IL}_{\mathrm{SM,STL}} \\
    &<
    1.643121 + 0.4 + 0.560000 + 0.560000 + 0.617101 \\
    &<
    3.780222~\mathrm{dB}
    \approx 3.78~\mathrm{dB}.
\end{align}

If the MM--MUX interface loss is neglected, the corresponding estimate
is
\begin{equation}
    \mathrm{IL}_{\mathrm{device,STL}}
    \approx 3.38~\mathrm{dB}.
\end{equation}

This estimate remains qualitative because the STL model was applied
only to the SM output taper. The MM taper, MUX, S-bend separator, and
full photonic lantern were not independently simulated using STL
geometries. The MM--MUX interface loss was estimated to be less than
$0.4~\mathrm{dB}$, while the S-bend loss was approximated by the MUX
loss because the two components were assumed to have comparable
cross-sectional geometries and aspect ratios. We were unable to simulate the stl MUX or S-bend separation components, or the full PL due to insufficient computational resources. Dedicated STL-based
simulations of the MM taper, MUX, S-bend separator, and complete
photonic lantern are therefore required before this value can be
treated as a quantitative prediction.

\section{FABRICATION}
\label{sec:fabrication}

The optimized photonic lantern was fabricated using a Nanoscribe
Photonic Professional GT2 TPP system. The structure was printed as a
standalone polymer-waveguide component.

The principal fabrication parameters were as follows:

\begin{itemize}
    \item Photoresist: IP-S resin.
    \item Substrate: $ 1 $-inch-diameter circular sapphire substrate.
    \item Objective: $ 63\times $, NA $ =1.4 $, immersion.
    \item Laser power: $ 40~\mathrm{mW} $.
    \item Nominal scan speed: $ 10~\mathrm{mm/s} $.
    \item Slicing distance: $ 100~\mathrm{nm} $.
    \item Hatching distance: $ 100~\mathrm{nm} $.
    \item Development procedure: PGMEA for 1~h, IPA for 1~h, and
    Novec 7100 for 1~h, followed by air drying.
\end{itemize}

For optimal resolution, the device was printed vertically. Because of
the aspect ratio of the finalized taper and MUX design, a support
structure was designed to hold the photonic lantern vertically. The
support structure contacted the device at a location along the
multimode taper selected to minimize optical loss. The contact region
consisted of eight $ 1~\mathrm{\mu m} $ contact points arranged around
the circumference of the multimode taper. The contact location was
chosen where the multimode taper had a diameter of
$ 9.429~\mathrm{\mu m} $, adapting the approach reported by Dana
et al.~\cite{Dana2024}. A  $ 1~\mathrm{\mu m} $ contact ring around this diameter was simulated, and the IL calculated was found to be negligible relative to overall design IL discussed previously, and thus is not considered further.

\begin{figure}[ht]
    \centering
    \includegraphics[width=\textwidth]{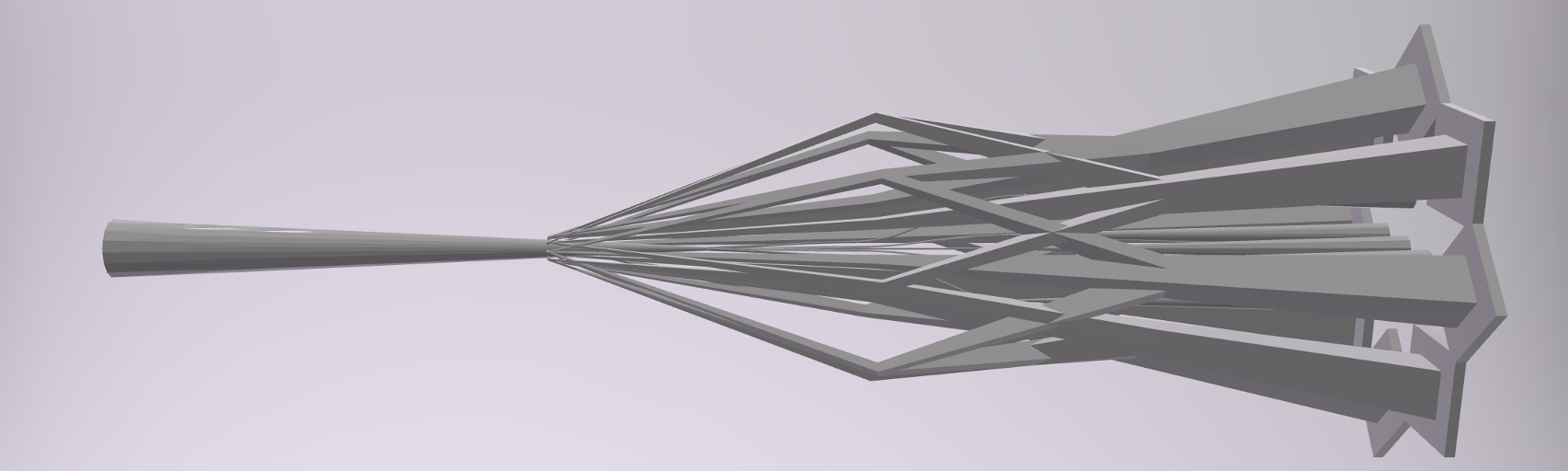}
    \caption{SolidWorks model of the photonic lantern and surrounding
    support structure.}
    \label{fig:support}
\end{figure}

\begin{figure}[ht]
    \centering
    \includegraphics[width=\textwidth]{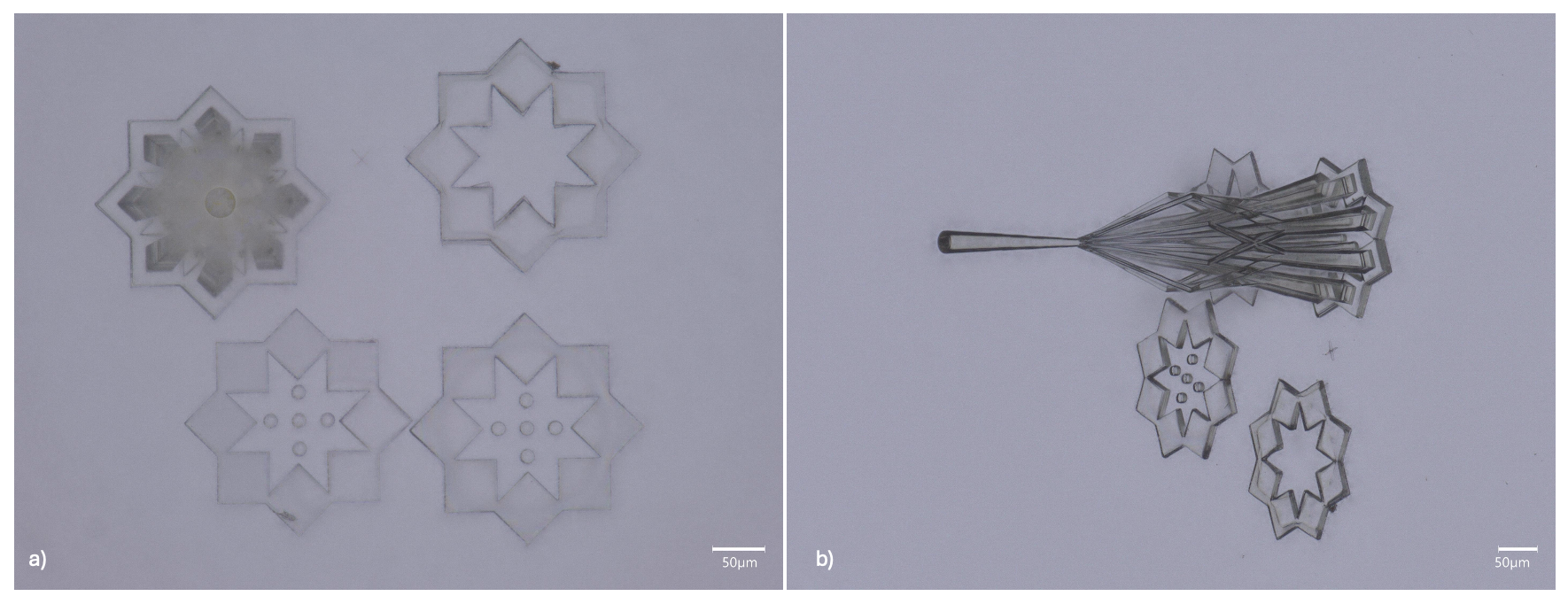}
    \caption{Optical-microscope images of the fabricated photonic
    lantern and support structure: (a) bottom view and (b) side view.
    The additional structures visible in both images are test bases
    fabricated alongside the device.}
    \label{fig:fabricated_device}
\end{figure}

Figure~\ref{fig:support} shows a 3D model of the support structure,
while Fig.~\ref{fig:fabricated_device} shows the completed device and
support structure.

\section{EXPERIMENTAL CHARACTERIZATION}
\label{sec:characterization}

\subsection{Free-space optical setup}

The fabricated lantern was characterized using a free-space optical
setup consisting of a helium-neon (HeNe) $ 635~\mathrm{nm} $ laser,
collimating and focusing optics, the photonic-lantern sample, and
camera-based imaging of the transmitted fields.

A 635nm laser was selected for initial characterization because its
visible wavelength simplified alignment and excitation of the
lantern. Future work will first reproduce these measurements using a
fiber-coupled HeNe source and will then replace that source with a
$ 1050~\mathrm{nm} $ laser to permit direct comparison with the
design-wavelength simulations.

\begin{figure}[ht]
    \centering
    \includegraphics[width=\textwidth]{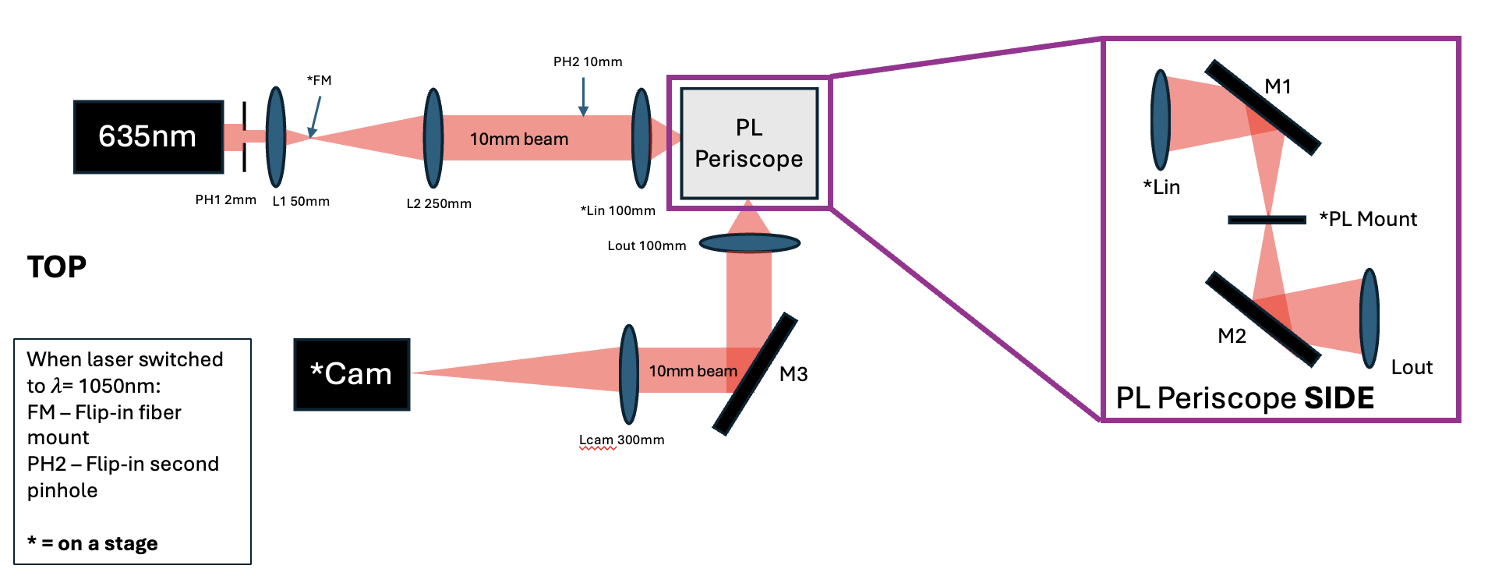}
    \caption{Free-space optical setup used for preliminary
    characterization of the photonic lantern.}
    \label{fig:optical_setup}
\end{figure}

As shown in Figure~\ref{fig:optical_setup}, the focusing and collimating optics immediately upstream and downstream of the PL Periscope (Lin and Lout) were designed to provide $f/10$, based on the MM and SM taper designs, as described in Section 2.1.
The camera optic, $L_{\mathrm{cam}}$, was selected to provide
Nyquist sampling of the focal spots with more than two pixels per
resolution element. For the $f/10$ single-mode output, the expected
spot size on the camera is
\begin{equation}
\frac{300}{10}\frac{0.635}{5.86} = 3.25
\ \mathrm{pixels/resel},
\end{equation}
where the camera pixel size is $5.86~\mathrm{\mu m}$.

\begin{figure}[ht]
    \centering
    \includegraphics[width=0.6\textwidth]{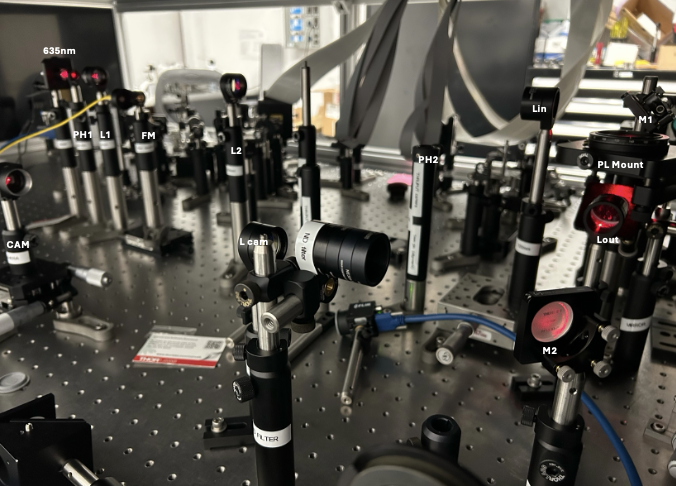}
    \caption{Benchtop image of optical test setup.}
    \label{fig:optical_setup2}
\end{figure}

Figure~\ref{fig:optical_setup} shows a schematic of the experimental
setup, and Figure~\ref{fig:optical_setup2} shows a picture of the benchtop setup. The lantern was printed vertically relative to the substrate,
with the single-mode ends located at the base of the device. The
periscope was designed to maintain the lantern in a vertical
orientation with respect to gravity, thereby minimizing mechanical
stress and reducing the risk of deformation or loss of structural
integrity during handling and operation.

The principal experimental parameters were as follows:

\begin{itemize}
    \item Thorlabs® CPS635R Laser Diode: $ 635~\mathrm{nm} $ collimated beam.
    \item PH1: $ 2~\mathrm{mm} $ pinhole used to set the beam size.
    \item Magnification optics: $ L_1=50~\mathrm{mm} $ and
    $ L_2=250~\mathrm{mm} $, providing $ 5\times $ magnification.
    \item Lantern focusing and collection optics:
    $ L_{\mathrm{in}}=100~\mathrm{mm} $ and
    $ L_{\mathrm{out}}=100~\mathrm{mm} $, providing $f/10$.
    \item PL Mount: PL was mounted in a 6 degree of freedom mount for precision alignment.
    \item Camera optic: $ L_{\mathrm{cam}}=300~\mathrm{mm} $, chosen to provide Nyquist sampling of focal spots.
    \item Flir camera: BFLY-U3-23S6M-C USB 3.0 Blackfly®, Monochrome Camera 
\end{itemize}

\subsection{Alignment procedure and optical results}

The system was initially aligned to the input plane of the photonic
lantern mount without the lantern substrate installed. Using
mechanical datums, each optical element was aligned with the optical
axis. A mirror was then inserted into the PL mount, and the tip and
tilt of the mount were adjusted using the back-reflected beam. A shear
plate was used to verify collimation in the appropriate sections of
the optical path. Finally, the camera position was adjusted to obtain
a well-focused point-spread function (PSF).

\begin{figure}[ht]
    \centering
    \includegraphics[width=0.6\textwidth]{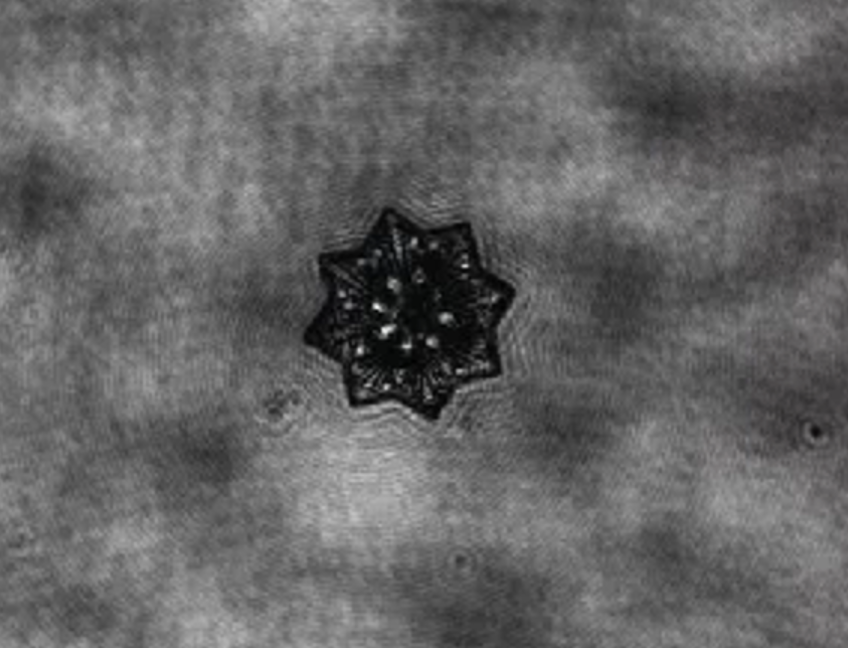}
    \caption{Image of the lantern with the $ L_{\mathrm{in}} $ optic
    removed. Illumination by the broad beam allowed the lantern to be
    centered in the transverse directions and the output plane to be
    brought into focus.}
    \label{fig:align}
\end{figure}

Following the initial alignment, the sapphire substrate containing the
PL was installed in the mount. Because the lantern was less than
$ 700~\mathrm{\mu m} $ long, direct visual alignment was impractical.
The $ L_{\mathrm{in}} $ optic was therefore removed, allowing the
lantern to be illuminated by a broad beam and its shadow to be
observed by the camera. This image was used to center the lantern in
the transverse $ x $ and $ y $ directions. The camera was then
refocused until the lantern output plane was sharply imaged, as shown
in Fig.~\ref{fig:align}.

The $ L_{\mathrm{in}} $ optic was subsequently reinserted into the
optical path, and was translated along the optical axis until
the lantern was positioned at the appropriate excitation plane.

\begin{figure}[ht]
    \centering
    \includegraphics[width=\textwidth]{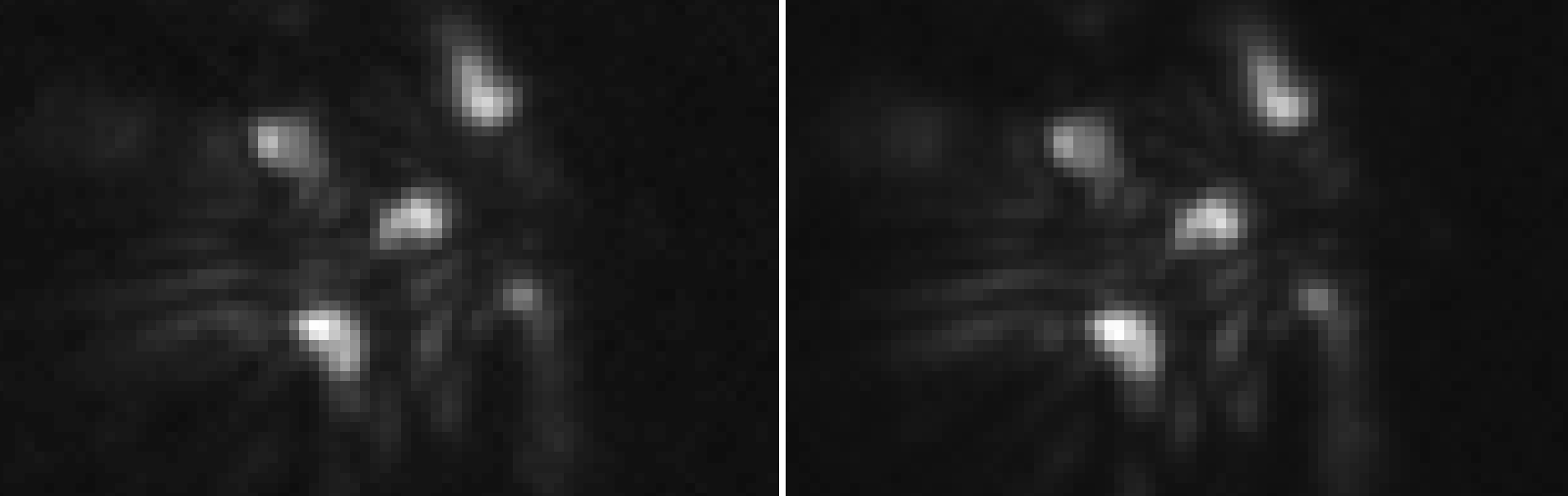}
    \caption{Camera images of the five photonic-lantern outputs
    following alignment and successful coupling.}
    \label{fig:outputs}
\end{figure}

After alignment, the input laser beam was successfully coupled into
the MM end of the PL. Light was guided through the lantern and
observed at each of the five SM output ports, as shown in Fig.~\ref{fig:outputs}.

\subsection{Photonic-lantern throughput estimate}

\begin{figure}[ht]
    \centering
    \includegraphics[width=\textwidth]{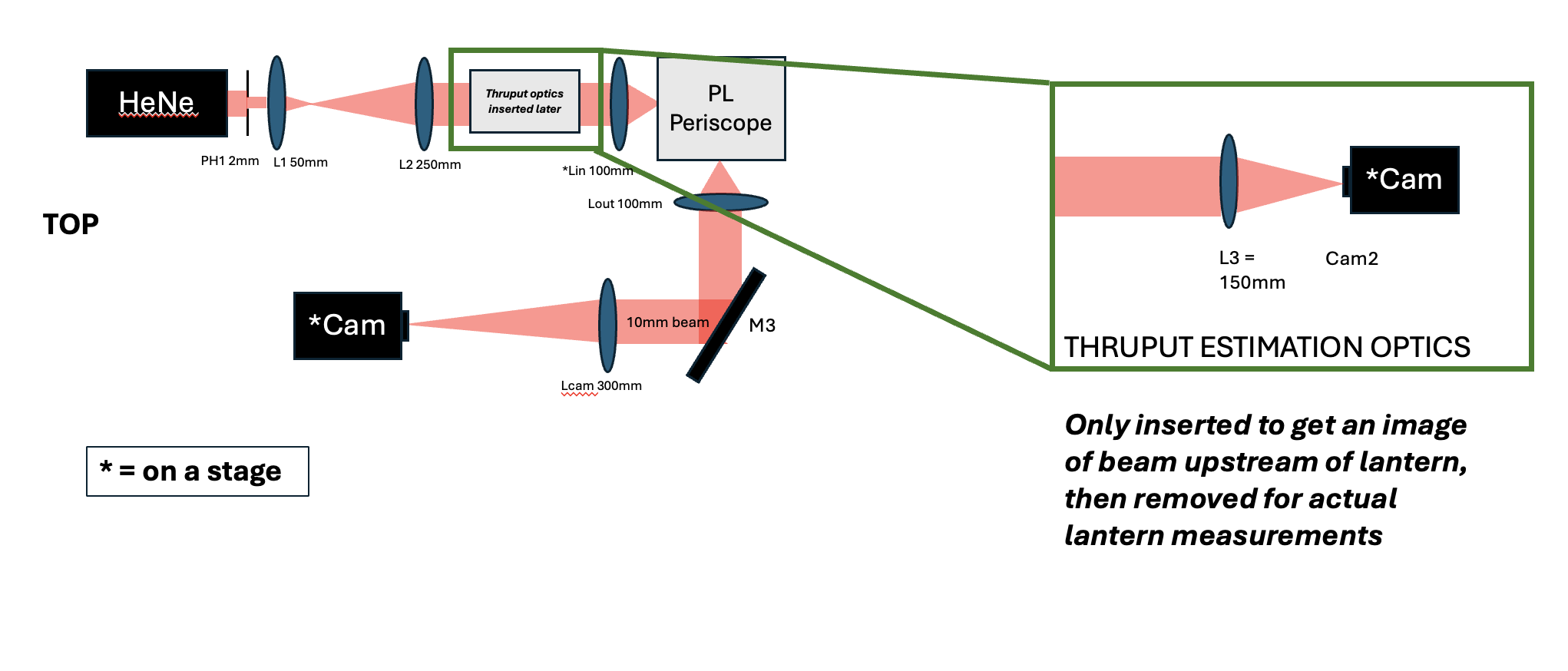}
    \caption{Optics inserted into the beam path to image the beam
    upstream of the photonic lantern for the reference measurement.}
    \label{fig:thruput_setup}
\end{figure}

The photonic-lantern throughput was estimated from camera images
acquired before and after the lantern. The optics used to collect the before image are shown in Fig.~\ref{fig:thruput_setup}. The five single-mode output
spots were enclosed within one combined region of interest (ROI).
Separate background images were recorded at the corresponding
exposure times.

The exposure times were
\begin{align}
    t_{\mathrm{before}}
    &=
    96.02~\mathrm{\mu s}, \\
    t_{\mathrm{after}}
    &=
    480.12~\mathrm{\mu s}.
\end{align}

No saturated pixels were present in either signal image. The maximum
pixel values were 151 before the lantern and 90 after the lantern,
compared with the 8-bit saturation value of 255.

For each measurement, the exposure-matched background image was first
subtracted:
\begin{equation}
    D_i(x,y)
    =
    I_i(x,y)-B_i(x,y),
\end{equation}
where $ I_i $ is the signal image, $ B_i $ is the corresponding
background image, and $ i $ denotes either the before or after
measurement. Signed pixel values were retained after subtraction.

\begin{figure}[ht]
    \centering
    \includegraphics[width=\textwidth]{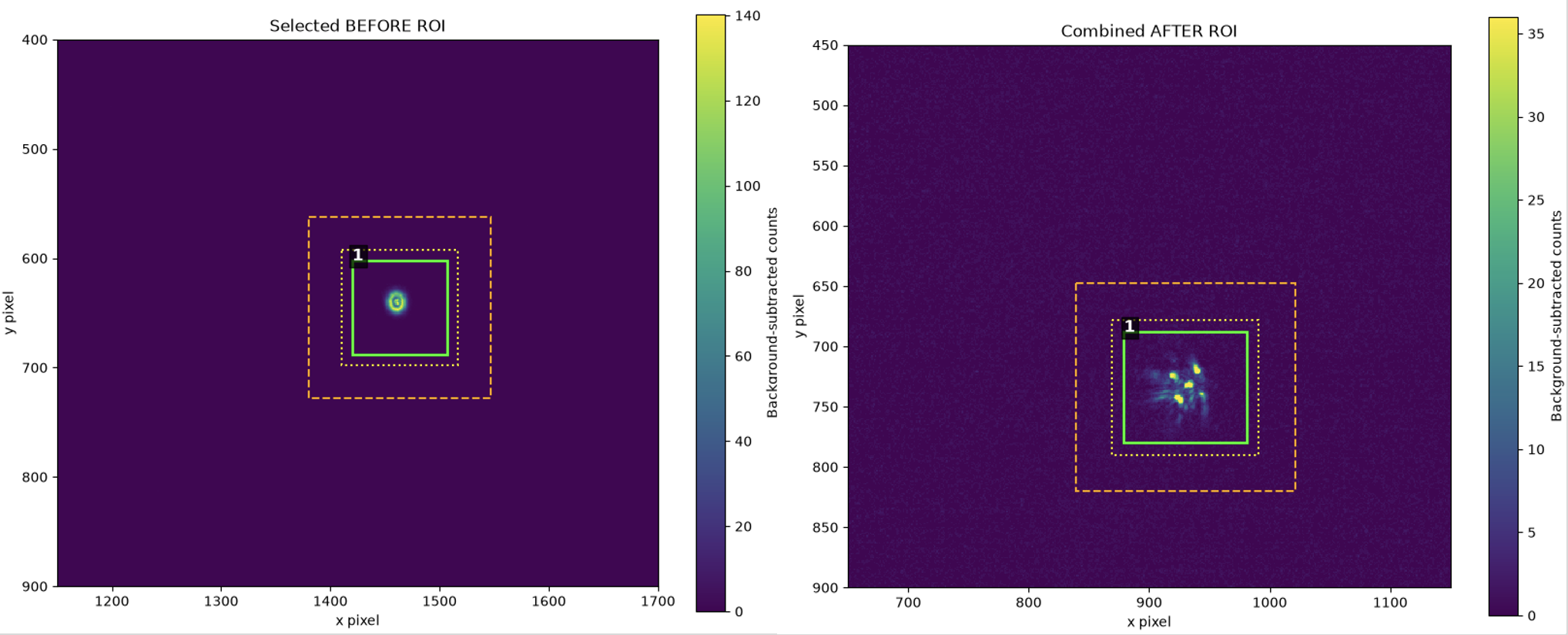}
    \caption{Background-subtracted images used to estimate throughput,
    with the selected regions of interest.}
    \label{fig:thruput_images}
\end{figure}

A residual local background level, $ b_i $, was then estimated using
a sigma-clipped mean of pixels in a rectangular ring surrounding the
signal ROI. A guard region separated the signal ROI from the
background ring to reduce contamination from the spot halo. The
corrected integrated signal was calculated as
\begin{equation}
    S_i
    =
    \sum_{(x,y)\in\mathrm{ROI}_i}
    \left[D_i(x,y)-b_i\right].
\end{equation}

The measured corrected sums were
\begin{align}
    S_{\mathrm{before}}
    &=
    33464.2~\mathrm{counts}, \\
    S_{\mathrm{after}}
    &=
    18974.4~\mathrm{counts}.
\end{align}

Because the two images were acquired with different exposure times,
the integrated signals were converted to count rates:
\begin{align}
    R_{\mathrm{before}}
    &=
    \frac{S_{\mathrm{before}}}{t_{\mathrm{before}}}
    =
    \frac{33464.2}{96.02}
    =
    348.513~\mathrm{counts/\mu s}, \\
    R_{\mathrm{after}}
    &=
    \frac{S_{\mathrm{after}}}{t_{\mathrm{after}}}
    =
    \frac{18974.4}{480.12}
    =
    39.5201~\mathrm{counts/\mu s}.
\end{align}

The estimated throughput was therefore
\begin{equation}
    T
    =
    \frac{R_{\mathrm{after}}}{R_{\mathrm{before}}}
    =
    \frac{S_{\mathrm{after}}/t_{\mathrm{after}}}
         {S_{\mathrm{before}}/t_{\mathrm{before}}}
    =
    0.113396.
\end{equation}

Thus, the measured throughput was
\begin{equation}
    \boxed{T\approx11.34\%}.
\end{equation}

An ROI-expansion test produced throughput values between approximately
$ 11.23\% $ and $ 11.36\% $ for expansion margins from 0 to 35
pixels. This plateau indicates that the primary spots and their
relevant halos were included. Larger expansion margins caused the
result to increase, likely because of residual stray light or spatial
background structure. Consequently, a representative preliminary
throughput is
\begin{equation}
    \boxed{T\approx11.3\%}.
\end{equation}

The corresponding insertion loss was calculated using
\begin{equation}
    \mathrm{IL}
    =
    -10\log_{10}(T).
\end{equation}

For $ T=0.113396 $,
\begin{equation}
    \mathrm{IL}
    =
    -10\log_{10}(0.113396)
    \approx
    9.45~\mathrm{dB}.
\end{equation}

Therefore, the measured result at the $ 635~\mathrm{nm} $ HeNe
wavelength can be summarized as
\begin{equation}
    \boxed{
        T\approx11.3\%,
        \qquad
        \mathrm{IL}\approx9.5~\mathrm{dB}
    }.
\end{equation}

This value represents the measured optical path. It corresponds
specifically to the lantern throughput only if losses and
collection-efficiency differences from intervening optics have been
excluded or calibrated.

\section{DISCUSSION}
\label{sec:discussion}

The free-space measurements provide preliminary experimental evidence
that light was successfully coupled into the MM input, guided through
the MUX, and delivered to all five output channels. At the
$ 635~\mathrm{nm} $ HeNe wavelength, the measured throughput was
approximately $ 11.3\% $, corresponding to an insertion loss of
approximately $ 9.45~\mathrm{dB} $. The stability of the result under
ROI expansion suggests that the primary output spots and their
associated halos were included in the analysis. The observation of
five distinct output spots confirms the basic guiding and fan-out
functionality of the fabricated lantern.

The component-level simulations at the design wavelength of
$1050~\mathrm{nm}$ produced an estimated total insertion loss of
$6.11~\mathrm{dB}$. This corresponds to a simulated transmission of
\begin{equation}
    T_{\mathrm{sim},1050}
    =
    10^{-6.11/10}
    \approx
    0.245,
\end{equation}
or approximately $24.5\%$. If the current experimental and simulated
values are placed side by side, the measured insertion loss is
approximately
\begin{equation}
    \Delta\mathrm{IL}
    =
    9.45-6.11
    =
    3.34~\mathrm{dB}
\end{equation}
greater than the nominal simulated estimate. However, this difference
should not be interpreted as a direct model-to-experiment discrepancy
because the measurement and simulation were performed at substantially
different wavelengths.

The wavelength difference is presently the most important limitation
on the comparison. The lantern geometry was optimized at
$ 1050~\mathrm{nm} $, whereas the initial experiment used a
$ 635~\mathrm{nm} $ HeNe laser to simplify visual alignment. At
$ 635~\mathrm{nm} $, the physical waveguide dimensions are
approximately $ 1050/635\approx1.66 $ times larger relative to the
wavelength than at the design wavelength. This changes the normalized
waveguide dimensions, modal effective indices, number of supported
modes, confinement, and adiabaticity of the taper and MUX transitions.
In particular, waveguides intended to be nominally single mode at
$ 1050~\mathrm{nm} $ may support additional modes at the shorter
wavelength. The refractive index and attenuation of the polymer can
also vary with wavelength. Consequently, the modal evolution and
insertion loss measured at $ 635~\mathrm{nm} $ are not expected to be
identical to those predicted at $ 1050~\mathrm{nm} $.

The simulated insertion loss also represents an average over selected
port-eigenmode launches, whereas the experiment used a focused,
coherent free-space beam. If the incident field excites a modal
coefficient vector $ \mathbf{a} $, the transmitted power depends on
the specific modal amplitudes and phases according to
\begin{equation}
    T(\mathbf{a})
    =
    \frac{
        \left\|\mathbf{H}\mathbf{a}\right\|^2
    }{
        \left\|\mathbf{a}\right\|^2
    }.
\end{equation}

The experimental throughput can therefore depend strongly on the beam
position, incidence angle, focal position, waist, phase-front
curvature, and polarization. A focused beam does not necessarily
excite the ten modeled input channels with equal power, and it is not
generally equivalent to the mean launch-dependent insertion loss
reported by the simulations. This sensitivity is particularly
important when the transmission eigenvalues are nonuniform.

The angular sensitivity of the photonic lantern provides an additional
consideration when interpreting the measured throughput. The simulated
far-field half-angles at the wide end of the taper, which enclose
approximately $95\%$ of the beam power, are listed in
Table~\ref{tab:far_field_angles}.

\begin{table}[ht]
\centering
\caption{Far-field half-angles for the five modes at the wide end of
the taper. The listed angles enclose approximately $95\%$ of the beam
power.}
\label{tab:far_field_angles}
\begin{tabular}{cc}
\hline
Mode & Half-angle (deg) \\
\hline
1 & 0.989 \\
2 & 1.526 \\
3 & 1.531 \\
4 & 1.896 \\
5 & 2.117 \\
\hline
\end{tabular}
\end{table}

These values can be compared with the angular precision of the
photonic-lantern tip--tilt alignment. For the most restrictive
half-angle, approximately $1^\circ$, a corresponding angular
misalignment would produce a lateral displacement between the
back-reflected and forward-propagating beams of approximately
\begin{equation}
\Delta x \simeq 2L\tan\theta
\approx 2(100~\mathrm{mm})\tan(1^\circ)
\approx 3.5~\mathrm{mm},
\end{equation}
for a card located approximately $100~\mathrm{mm}$ upstream of
$L_{\mathrm{in}}$. For small angles, this can be estimated as
$\Delta x \approx 2L\theta$, with $\theta$ expressed in radians.
So even modest angular offsets can produce measurable beam
displacements and may make precise co-alignment difficult. The
measured throughput can therefore reflect not only the intrinsic
launch-dependent insertion loss, but also the sensitivity of the
lantern to beam position and angular alignment. Future measurements
could improve this alignment by isolating the back-reflected beam,
for example using a beam splitter or beam cube.

Several numerical assumptions may also contribute to differences
between the predicted and measured performance. The MM taper, MUX,
and SM output taper were simulated independently, and their insertion
losses were subsequently added in decibels. This approach assumes
independent cascaded components with negligible interface reflections,
interference, and additional mode mismatch. The assembled device may
contain coherent coupling and reflections at the taper--MUX
interfaces that are not represented by this approximation. A
continuous end-to-end simulation, or a consistent cascade of the
complex component S-matrices, would provide a more complete prediction
of the assembled-device response.

The discretized cylindrical geometry may introduce additional
numerical discontinuities. This interpretation is supported by the
reduction in simulated SM-taper insertion loss from
$ 1.98347~\mathrm{dB} $ to $ 0.617101~\mathrm{dB} $ when the taper
was represented using an STL model. The corresponding estimate of
$ 3.780222~\mathrm{dB} $ for the complete STL-based device would represent
a transmission of approximately
\begin{equation}
    T_{\mathrm{device,STL}}
    =
    10^{-3.780222/10}
    \approx
    0.45,
\end{equation}
or approximately $ 52.2\% $. However, this device-level value assumes
that the MM taper experiences the same reduction in loss as the SM
taper. It should therefore be regarded as a qualitative projection
rather than a validated performance prediction. A dedicated
STL-based simulation of the MM taper and, ultimately, the complete
lantern is required before this estimate can be compared
quantitatively with experiment.

Differences between the nominal design and fabricated structure may
produce additional loss. Relevant fabrication effects include
polymerization shrinkage, finite voxel dimensions, slicing and
hatching discretization, surface roughness, incomplete development,
and deviations in the taper diameter, arm spacing, MUX trajectory,
and total device length. Small changes in the
$ 1.0~\mathrm{\mu m} $ MUX waveguide diameter can significantly alter
modal confinement because the dimensions are comparable to the design
wavelength. Bending, tilt, or deformation of the vertically printed
structure may also modify the coupling between adjacent waveguides.

The simulations assumed a real and spatially uniform refractive index of
$n=1.53$. This excludes polymer absorption, surface scattering,
refractive-index variations caused by the exposure dose, and material
dispersion. Facet roughness and Fresnel reflections at polymer--air or
polymer-substrate interfaces may contribute additional experimental
loss. For normal incidence, the Fresnel power reflectance at an
interface between materials with refractive indices $n_1$ and $n_2$ is
\begin{equation}
R = \left(\frac{n_1-n_2}{n_1+n_2}\right)^2.
\end{equation}
For $n_1=1.53$ and air with $n_2=1.00$, this gives
$R\approx0.0439$, or approximately $4.4\%$ loss per polymer-air
surface. Similarly for the, polymer-sapphire transition, where sapphire has an index of $n_1=1.7659$, would appoximately give $R\approx0.00512$, or $0.5\%$ loss.
And finally, the sapphire-air surface would approximately give $R\approx0.0758$, or $7.5\%$ loss.
Summing as$(1-0.044)*(1-0.005)*(1-0.07)$, would give a total of about $11.5\%$ loss, or an additional $0.53$ dB.

These estimates assume normal incidence and neglect polarization,
surface roughness, and scattering. Measuring the refractive index and
attenuation of the processed material at both $635~\mathrm{nm}$
and $1050~\mathrm{nm}$, together with characterizing the relevant
surface conditions, would reduce these uncertainties.

The measured value may also contain losses that are external to the
lantern. The count-rate normalization assumes that the camera response
is linear with exposure time and that the detector gain remains
constant. Although no saturated pixels were observed, camera
linearity should be verified over the exposure range used. Differences
in the optical paths of the before- and after-lantern measurements may
include lens transmission, substrate transmission, Fresnel reflection,
collection numerical aperture, aberrations, and detector-response
effects. Power emitted outside the collection numerical aperture or
outside the selected ROI would be interpreted as lantern loss, while
stray light inside the ROI could increase the estimated throughput.
The ROI-expansion plateau constrains one part of this uncertainty but
does not account for these systematic effects.

The relative intensities of the five output spots may contain
information about the modal content of the incident field. However,
unequal output powers cannot yet be assigned uniquely to intrinsic
device nonuniformity because they can also result from input
misalignment, polarization, or excitation of a particular coherent
modal superposition. Furthermore, the camera images provide intensity
but not phase information. A quantitative comparison of the output
distribution will require simulations at the experimental wavelength
using the measured input beam and a consistent model of the
collection optics.

The next experimental step is therefore to repeat the throughput
measurement at the design wavelength of $ 1050~\mathrm{nm} $ using
wavelength-appropriate optics and a calibrated detector. The reference
and lantern measurements should use the same optical path wherever
possible, with the transmissions of unavoidable additional components
measured independently. Position, focus, tip, tilt, and polarization
scans should be performed to determine the optimum coupling condition
and quantify alignment sensitivity. Repeated measurements on multiple
devices would allow fabrication repeatability and measurement
uncertainty to be established.

In parallel, the device should be simulated at $ 635~\mathrm{nm} $
using wavelength-dependent material properties, the complete guided
mode set, and renewed cell- and modal-basis convergence studies. A
wavelength sweep between $ 635~\mathrm{nm} $ and
$ 1050~\mathrm{nm} $ would help distinguish spectral behavior from
fabrication and alignment losses. Measurements of the fabricated
dimensions could then be incorporated into an as-built model.
Together, measurements at $ 1050~\mathrm{nm} $ and simulations at
$ 635~\mathrm{nm} $ will provide the two direct comparisons needed to
determine whether the remaining loss is primarily caused by
wavelength-dependent modal behavior, fabrication deviations,
numerical approximations, or the experimental optical path.

Strong wavelength dependence was also observed in a preliminary
simulation of the MUX alone. When the simulation wavelength was
changed from the optimized value of $ 1050~\mathrm{nm} $ to the
experimental wavelength of $ 635~\mathrm{nm} $, while retaining the
same nominal MUX geometry, the simulated insertion loss increased from
$ 0.56~\mathrm{dB} $ to $ 9.74~\mathrm{dB} $. The corresponding
output-power nonuniformity, $ \Delta_{\mathrm{out}} $, increased from
$ 0.86~\mathrm{dB} $ to $ 7.37~\mathrm{dB} $. Thus, the wavelength
change produced increases of approximately $ 9.18~\mathrm{dB} $ in
insertion loss and $ 6.51~\mathrm{dB} $ in output-power nonuniformity.

The insertion-loss values correspond to a reduction in simulated MUX
transmission from approximately
\begin{equation}
    T_{\mathrm{MUX},1050}
    =
    10^{-0.56/10}
    \approx
    0.879
\end{equation}
at $ 1050~\mathrm{nm} $ to
\begin{equation}
    T_{\mathrm{MUX},635}
    =
    10^{-9.74/10}
    \approx
    0.106
\end{equation}
at $ 635~\mathrm{nm} $. The substantially larger
$ \Delta_{\mathrm{out}} $ at $ 635~\mathrm{nm} $ further indicates
that the shorter-wavelength response is not only more lossy but also
considerably less uniform across the modeled modal channels.

These values describe row-summed output-power nonuniformity and should
not be interpreted as conventional MDL. Conventional MDL requires the
eigenvalues of the complex transmission matrix
$ \mathbf{H}^{\dagger}\mathbf{H} $.

This preliminary comparison supports the expectation that operation
away from the design wavelength can substantially degrade MUX
performance. Nevertheless, the $ 635~\mathrm{nm} $ result should be
verified using wavelength-dependent material properties and renewed
convergence studies for the EME cell count, modal-basis size, port
modes, and transverse simulation domain. The numerical similarity
between the simulated MUX loss of $ 9.74~\mathrm{dB} $ and the
measured complete-path loss of approximately $ 9.45~\mathrm{dB} $
should not be interpreted as direct agreement. The former describes
only the simulated MUX, whereas the latter includes the complete
fabricated lantern and potentially the surrounding experimental
optical path. Direct validation will require both a complete-device
simulation at $ 635~\mathrm{nm} $ and an experimental measurement at
the optimized wavelength of $ 1050~\mathrm{nm} $.

\section{CONCLUSION}
\label{sec:conclusion}

A compact $ 1\times5 $ photonic lantern optimized for operation at
$ 1050~\mathrm{nm} $ was designed, simulated, fabricated by
two-photon polymerization, and characterized using a free-space
optical system. Component-level EME simulations at the design
wavelength predicted a total insertion loss of approximately
$ 6.11~\mathrm{dB} $ under the assumption that the losses of the MM
taper, MUX, and one SM output taper can be added independently. An
STL-based representation reduced the simulated SM-taper loss from
$ 1.98347~\mathrm{dB} $ to $ 0.617101~\mathrm{dB} $, indicating that
a continuous geometric model may better represent the fabricated
transition. Extrapolation of this improvement to the MM taper produced
a qualitative full-device loss estimate of approximately
$ 3.78~\mathrm{dB} $, although this value requires validation through
a dedicated STL-based MM-taper simulation.

Preliminary characterization at $ 635~\mathrm{nm} $ demonstrated
successful coupling through the lantern and observable light at all
five output ports. The measured throughput was approximately
$ 11.3\% $, corresponding to an insertion loss of approximately
$ 9.45~\mathrm{dB} $. This measurement cannot be compared directly
with the $ 1050~\mathrm{nm} $ device-level simulation because the
lantern was operated substantially outside its optimized wavelength.
A separate MUX-only simulation supports this interpretation. Changing
the simulation wavelength from $ 1050~\mathrm{nm} $ to
$ 635~\mathrm{nm} $ increased the MUX insertion loss from
$ 0.56~\mathrm{dB} $ to $ 9.74~\mathrm{dB} $ and increased the
output-power nonuniformity, $ \Delta_{\mathrm{out}} $, from
$ 0.86~\mathrm{dB} $ to $ 7.37~\mathrm{dB} $. These results indicate
strong wavelength dependence and suggest that the preliminary
measured loss may be dominated in part by operation away from the
design wavelength.

Future work will include calibrated throughput measurements at
$ 1050~\mathrm{nm} $, complete-device simulations at
$ 635~\mathrm{nm} $, and wavelength sweeps incorporating dispersive
material properties. Additional studies will examine alignment and
polarization sensitivity, modal excitation, fabrication tolerances,
and as-built device geometry. Continuous STL-based simulations of the
MM taper and complete lantern will also be performed. Conventional
MDL will be calculated from the eigenvalues of
$ \mathbf{H}^{\dagger}\mathbf{H} $ and reported separately from the
row-summed output-power nonuniformity used in the present analysis.
These measurements and simulations will enable a direct assessment of
the design and clarify the relative contributions of wavelength
mismatch, numerical discretization, fabrication deviations, and
experimental coupling losses. The present results nevertheless
demonstrate successful fabrication and light guiding in a standalone
additively manufactured photonic lantern, supporting its continued
development for free-space-coupled astrophotonic applications.

\section*{DISCLOSURES}

The authors declare no conflicts of interest.

\acknowledgments

This work was performed under the auspices of the U.S. Department of
Energy by Lawrence Livermore National Laboratory under Contract
DE-AC52-07NA27344. This document number is LLNL-CONF-2024259.

\end{document}